# Managing Cloud networking costs for data-intensive applications by provisioning dedicated network links

Igor Sfiligoi
University of California San Diego
La Jolla, CA, USA
isfiligoi@sdsc.edu

Michael Hare
University of Wisconsin–Madison
Madison, WI, USA
michael.hare@wisc.edu

David Schultz
University of Wisconsin–Madison
Madison, WI, USA
david.schultz@icecube.wisc.edu

Frank Würthwein
University of California San Diego
La Jolla, CA, USA
fkw@ucsd.edu

Benedikt Riedel
University of Wisconsin–Madison
Madison, WI, USA
briedel@icecube.wisc.edu

Tom Hutton
University of California San Diego
La Jolla, CA, USA
hutton@sdsc.edu

Steve Barnet
University of Wisconsin–Madison
Madison WI USA
barnet@icecube.wisc.edu

Vladimir Brik
University of Wisconsin–Madison
Madison WI USA
vbrik@icecube.wisc.edu

*Abstract*—Many scientific high-throughput applications can benefit from the elastic nature of Cloud resources, especially when there is a need to reduce time to completion. Cost considerations are usually a major issue in such endeavors, with networking often a major component; for data-intensive applications, egress networking costs can exceed the compute costs. Dedicated network links provide a way to lower the networking costs, but they do add complexity. In this paper we provide a description of a 100 fp32 PFLOPS Cloud burst in support of IceCube production compute, that used Internet2 Cloud Connect service to provision several logically-dedicated network links from the three major Cloud providers, namely Amazon Web Services, Microsoft Azure and Google Cloud Platform, that in aggregate enabled approximately 100 Gbps egress capability to on-prem storage. It provides technical details about the provisioning process, the benefits and limitations of such a setup and an analysis of the costs incurred.

## I. INTRODUCTION

Scientific high throughput computing (HTC) needs typically vary with time and most fields occasionally experience significant spikes in demand, e.g., right before major conferences. Since provisioning dedicated on-prem resources for peak demand is prohibitively expensive, multi-domain on-prem research platforms, like the Open Science Grid (OSG) [1], have seen significant success in providing additional resources to scientists in times of need, by either borrowing compute capacity from unrelated science domains or by abstracting access to specialized resources, like XSEDE HPC centers [2]. The amount of available on-prem spare capacity is however limited, and commercial Cloud resources can provide significant additional capacity on very short notice [3-5].

Unlike on-prem capacity bursting, where capacity limits are mostly dominated by administrative burdens, e.g., applying for XSEDE allocations, commercial Cloud bursting is mostly dominated by budget considerations. None of the major Cloud providers, namely Amazon Web Services (AWS), Microsoft Azure and Google Cloud Platform (GCP), limit what can be done on their resources (within reason); as long as customers are willing and able to pay, they are happy to keep providing the requested resources. Unlike most on-prem resources, however, users get charged for most services, including compute, storage and networking, with some service types more cost effective than others.

Past papers have explored the compute instance cost-effectiveness and the feasibility of using only the most cost-effective ones [5,6], with encouraging results. We are however not aware of any work that addresses the use of cost-effective network options in support of data-intensive applications in large setups.

The major networking cost in the three major Clouds is egress traffic, e.g., data produced on Cloud compute instances being moved back to on-prem storage. All other network traffic is either free or has negligible cost, to the first approximation. At the time of writing, the most cost-effective networking option is the use of dedicated network links; depending on the mode of use, it reduces the egress costs for data-intensive applications between 50% and 75%.

In this paper we describe the experience of running a half-day Cloud burst in support of IceCube production compute, that required a peak egress network traffic of over 10 GigaBytes per second (GBps) to upload results of simulation compute generated using about 100 fp32 FLOPS of Cloud compute power. All egress traffic was routed over dedicated links, which were provisioned using Internet2 Cloud Connect service [7,8].

Sections 2 provides an overview of the IceCube science and the specific application being run. Section 3 provides an overview of the HTC setup used to execute the Cloud burst. Section 4 provides the description of the steps needed to provision the dedicated network links. Section 5 provides the description of the Cloud burst, with an emphasis on the data movement part of the exercise. Finally, section 6 provides an analysis of the costs incurred during the exercise.

This work was partially funded by the US National Science Foundation (NSF) through grants OAC-1941481, MPS-1148698, OAC-1841530, OAC-1826967 and OPP-1600823.

Pre-print version, Apr. 2021.

## II. IceCube science and application

The IceCube Neutrino Observatory [9] is the world's premier facility to detect neutrinos with energies above 1 TeV and an essential part of multi-messenger astrophysics. IceCube is composed of 5160 digital optical modules (DOMs) buried deep in glacial ice at the geographical south pole. Neutrinos that interact close to or inside of IceCube produce secondary particles, often a muon. Such secondary particles produce Cherenkov (blue as seen by humans) light as they travel through the highly transparent ice. Cherenkov photons detected by DOMs can be used to reconstruct the direction and energy of the parent neutrino.

IceCube is built into a naturally existing medium, i.e., glacial ice. There was *a priori* only limited information regarding the optical properties of the instrumented ice. Because of this, a significant amount of simulation data is needed to properly calibrate the employed instruments. The optical properties of the glacial ice greatly affect the pointing resolution of IceCube. Improving the pointing resolution has two effects in this case: greater chance to detect astrophysical neutrinos and better information sent to the community. While IceCube can detect all flavors and interaction channels of neutrinos, about two-thirds of the flux reaching IceCube will generate a detection pattern with a large angular error, which is mostly driven by systematic effects. Similarly, different optical models have a great effect on the reconstructed location of an event on the sky. The comparatively minute field of view of partner observatories and telescopes requires IceCube to provide information as accurate as possible, as outlined in Fig. 1.

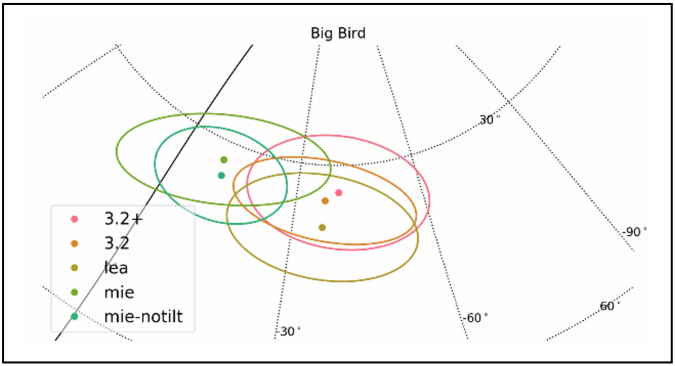

Fig. 1. Impact of the IceCube detector calibration on science results. Pointing area based on different versions of the detector calibration.

The most computationally intensive part of the IceCube simulation workflow is a photon propagation code, a.k.a. ray-tracing, and that code can greatly benefit from running on GPUs [10]. IceCube requires about 40 fp32 EFLOP-hours of compute each year to produce enough simulation to properly understand the detector. The application is high throughput in nature, with each photon simulation being independent of the others. Different variants of simulation have different egress data to compute ratios, ranging from 40 to 800 MegaBytes (MB) per fp32 TFLOP-hour of compute.

The workflow running the above application simulates a fixed number of photons, writes the resulting input to temporary local storage, and finally uploads that file to central storage after the compute is completed. This keeps bookkeeping simple and helps with error management. A characteristic of this approach is the very spiky nature of network traffic; most of the network traffic happens in a very narrow window just before job termination.

## III. HTCondor as HTC workload manager

IceCube's production setup uses HTCondor as the batch system. Most of the time, their compute resources come partially from local on-prem infrastructure and partially from remote systems, dynamically provisioned through the Open Science Grid (OSG) [11]. Extending the provisioning to Cloud resources is thus just a minor operational change.

IceCube normally does not run on Cloud resources, so it does not have a permanent provisioning infrastructure in place. We had however run a few other Cloud experiments in the past year [4,6], so we used a very similar provisioning setup this time, too. After creating the base virtual machine (VM) images using the standard OSG-provided worker node software, the actual large-scale provisioning was delegated to native group provisioning mechanisms, namely Spot Fleets on AWS, VM scale sets (VMSS) on Azure and Instance Groups on GCP. While the three Cloud providers use different implementations, the operational semantics is quite similar among the three. It should also be noted that each region in each Cloud provider is essentially independent, so we had to set up and operate this infrastructure in a dozen independent environments.

HTCondor manages only the compute part of the workflow, scheduling jobs to compute resources and dealing with error handling. The data movement to and from the central on-prem storage is handled independently by the jobs themselves; only log files useful for progress monitoring and debugging are handled explicitly by HTCondor. All IceCube applications are wrapped within a script that provides data movement functionality using GridFTP as a protocol.

## IV. Dedicated network links

For many users, networking is a black box. They likely know the IP, or DNS name, of the target location; however, how data moves to and from the compute node and the destination is completely opaque. This is actually a design goal of the Internet and has proven to be very successful in practical terms.

Most of the time, there is indeed no good reason for the final users to ponder about the details of network routing in and out of the Clouds; its opaque nature allows for both flexibility and optimization at the infrastructure level. However, when trying to optimize network costs, choosing the right route is essential. Data leaving a Cloud region over regular routes is charged at a much higher rate than data routed over dedicated links. On the flip side, there are no fixed costs for using the regular routes, but all major Cloud providers charge a fixed hourly fee for dedicated links, in addition or in lieu of data transfer fees. Table I provides a summary of the prices charged by AWS, Azure and GCP for their network services through a service provide like Internet2, as of December 2020. The reader should also be aware that dedicated links have a fixed max throughput associated with them, expressed in Gigabits per second (Gbps), while the default routes have no such restrictions [12].

TABLE I. SUMMARY OF NETWORK EGRESS COSTS IN THE USA

| Route, max. throughput | Fixed costs (per hour) | Egress costs (per TB) |
|---|---|---|
| Default route, unlimited | $0.00 | $80-$85 |
| Dedicated, 2 Gbps | $0.57-$1.19 | $20-$25 |
| Dedicated, 5 Gbps | $1.25-$2.99 | $20-$25 |
| Dedicated, 10 Gbps | $2.36-$4.65 | $20-$25 |
| Dedicated unmetered, 5 Gbps | $35 | $0.00 |

Provisioning dedicated links to local on-prem resources is however not straightforward. The major difference compared to provisioning compute resources in the Clouds is the fact that it involves three independent parties: the Cloud provider, the intermediate network provider, and the local on-prem networking team. In most circumstances, the final user does not have access to the non-Cloud layers; it thus becomes necessary to establish a relationship with the on-prem networking personnel, which in turn may need to collaborate with the intermediate network provider's personnel. Human interaction thus plays a significant part in establishing and tearing down of dedicated links. It also makes full automation of the provisioning process virtually impossible in most circumstances.

A dedicated network link also requires the allocation of a dedicated IP address range that is routable in the on-prem networking environment, with at least one IP address per compute resource being provisioned in the Cloud. While private IP addresses are acceptable, this typically requires some additional planning and coordination at the on-prem campus level, since that range cannot overlap with any other IP address range already in use on-prem.

For our exercise, we had two on-prem campuses to which egress data was flowing: University of Wisconsin–Madison (UW) and University of California San Diego (UCSD). UW is IceCube's home institution and the host of IceCube's central storage, so most of the egress data flowed there. UW's research network link is however limited to 100 Gbps, which would be shared with other users, so UCSD was added as a secondary egress destination to allow us to have at least 100 Gbps of available on-prem network bandwidth.

The dedicated links were provisioned using Internet2's Cloud Connect service, which acts as a peering provider for all three Cloud services, namely AWS Direct Connect, Microsoft Azure ExpressRoute and Google Cloud Interconnect. This allows US research institutions to provision the dedicated links using software defined networking (SDN) techniques, without any changes in their physical infrastructure. UW chose to provision a set of BGP-based Layer 3 virtual private networks (L3VPNs) to Internet2 via their regional aggregator, BTAA OmniPop. UCSD first provisioned a Layer 2 virtual private network (L2VPN) over their regional provider, CENIC, and then layered on top a BGP-based L3VPN with Internet2.

### A. Provisioning procedure

The provisioning procedure is very different for the three Cloud providers. A dedicated network link is the easiest to provision in GCP. The final user has to perform just four steps:

- Create a GCP virtual private cloud (VPC) instance associated with the chosen IP address range.
- Create subnets to associate with different Cloud zones.
- Create a GCP Cloud Router.
- Create the Google Cloud Interconnect.

The last step returns a key string that has to be shared with the on-prem networking personnel, which will in turn use it to establish the proper routing though Open Exchange Software Suite (OESS), a software-defined exchange (SDX) [13]. The provisioned bandwidth for the link is also selected in the OESS.

Provisioning in Microsoft Azure is conceptually very similar from the user's point of view, although the elements involved are slightly different:

- Create an Azure virtual network (VN) instance associated with the chosen IP address range.
- Create a gateway subnet for the Cloud Router to use, alongside a main subnet that will be used by the compute instances.
- Create the Azure ExpressRoute (ER) instance; this returns a key string that the on-prem networking personnel can use with OESS. The provisioned bandwidth and the billing option are selected by the final user during this step; Azure was the only provider used that allowed to choose between metered and unmetered billing.
- Create an Azure virtual network gateway (VNG). Care should be taken to select the appropriate variant (SKU), as only premium SKUs support throughputs of over 1 Gbps.
- Create an Azure network connection object, linking the ER to the VNG.

As with GCP, the on-prem networking personnel uses the provided key string to establish the routing to Azure via OESS. One peculiarity of Azure ExpressRoute is the fact that it uses the same IP address for routing all links provisioned in a specific peering point, which makes it impossible to use a single VPN to handle more than one dedicated link. This required us to create several independent L3VPNs in order to work around that.

Finally, AWS uses a completely different provisioning procedure. The most notable difference is that the provisioning request is initiated by the on-prem networking personnel, again using OESS, not by the final user; the provisioned bandwidth is specified during that step. After that is done, the number of steps the final user has to perform is also significantly longer:

- Accept the AWS Direct Connect request.
- Create an AWS virtual private cloud (VPC) instance associated with the chosen IP address range.

- Create subnets to associate with different Cloud zones.
- Create an AWS VPC Internet router.
- Create an AWS virtual private gateway (VPG).
- Associate VPG to VPC.
- Create a Direct Connect Gateway (DCG).
- Create an AWS Virtual Interface (VIF); this returns a BGP key and gateway IP addresses that must be conveyed to the on-prem networking personnel to finalize the routing configuration.
- Configure the VPC routing to use the VPG for on-prem traffic, and the Internet router for all other destinations.
- Associate DCG to the VPG.

Deprovisioning also varies between the three Cloud providers. In order to stop billing, GCP requires only the destruction of the Cloud Router. Similarly, on AWS the user just needs to delete the VIF. On Azure, the de-provisioning has to be initiated by the on-prem networking personnel, and billing continues until both the on-prem and the Cloud user delete the ExpressRoute objects. Of course, for a complete cleanup all the above steps need to be undone in all the Clouds.

### B. The need for many dedicated network links

Internet2 has network peering points with the three Cloud providers in several US cities. As of November 2020, most locations have only a few 10 Gbps connections into each Cloud provider, with the notable exception of 100 Gbps links into Azure in Silicon Valley. All of the links are shared, so no single customer is allowed to get the full bandwidth of any link. The reservable network bandwidth for a dedicated link is also quite rigid; one cannot request any arbitrary value, but has to pick among a pre-determined set, which happens to be 10, 5, and 2 Gbps at the high end.

All Cloud providers tie a provisioned dedicated network path with a specific Cloud compute region; AWS and GCP have additional restrictions, pre-assigning each Cloud compute region to a specific Internet2 peering location. Additional intra-Cloud routing is possible in all the Cloud providers, but comes at an additional cost, which is typically comparable to the on-prem egress cost; we thus avoided that option unless strictly necessary, i.e. when the used Cloud compute region did not have a feasible Internet2 peering location, i.e. AWS Oregon and GCP US East 1.

Using multiple dedicated network links was thus a necessity; we needed at least one link per Cloud compute region of each provider. Moreover, since we expected to provision enough compute capacity to egress over 10 Gbps in several of the largest Cloud regions, we had to provision multiple dedicated links for each of those regions. Indeed, we provisioned all the available peering capacity that could be provisioned using 2 and 5 Gbps links that Internet2 had with AWS and GCP in early November 2020, although we did not need all of Azure peering capacity. The complete list of the 21 dedicated links provisioned for the exercise is available in Table II, all of which used the metered billing model. You may notice that we provisioned two links that connected to Cloud regions outside the USA; while Azure does charge a higher hourly cost, the metered egress costs are identical. It thus made sense to use the additional compute capacity to better utilize the available network bandwidth.

TABLE II. PROVISIONED DEDICATED NETWORK LINKS

| Cloud Provider | On-prem storage | Peering location | Compute region | Max. throughput (Gbps) |
|---|---|---|---|---|
| AWS | UW | Dallas, TX | N. Virginia | 5, 5, 2, 2 |
| AWS | UW | Chicago, IL | Ohio | 5, 2, 2 |
| AWS | UW | San Jose, CA | California | 5 |
| AWS | UCSD | San Jose, CA | California | 5 |
| Azure | UW | Silicon Valley | East US | 10, 10 |
| Azure | UW | Silicon Valley | South Central US | 10 |
| Azure | UW | Silicon Valley | West EU | 10 |
| Azure | UW | Chicago, IL | UK South | 2 |
| Azure | UCSD | Silicon Valley | West US | 10 |
| GCP | UW | US Central | US Central | 5, 5, 2 |
| GCP | UW | US West 1 | US West 1 | 5, 5 |
| GCP | UW | US East 4 | US East 4 | 5 |

Note that for the Cloud regions where we provisioned more than one link, we partitioned the provisioned compute resources among them, i.e., each compute resource was assigned to exactly one dedicated network link.

## V. EXECUTING THE CLOUD BURST

As mentioned in Section 3, we had extensive experience executing Cloud bursts [4, 6], so we did not expect any problems with the compute capacity provisioning. The only major uncertainty was the available amount of compute in the various regions; while we had the measured values from the previous runs, we expected that the COVID-19 pandemic has since changed the global usage patterns, and thus the Cloud spare capacity. Having reliable estimates for the Cloud compute capacity is important for dedicated network provisioning. If we provisioned too much network capacity, we would have wasted money on underutilized links. If we provisioned too little, we could not have made full use of all the available compute capacity, or worse, we could have saturated the network links and wasted compute resources while waiting for network transfers to complete. Given the uncertainties, we ended up with outliers on both ends.

Apart from capacity planning, the used workload also made for a challenging setup. As mentioned in Section 2, the IceCube photon propagation simulation has a very spiky network transfer pattern; virtually all of the traffic happens just before the job termination. In order to both maximize the provisioned network bandwidth and minimize idle compute capacity, one thus has to randomize the job runtime as much as possible; in the case of dynamically provisioned compute resources, this means a moderately slow provisioning pace.

The chosen IceCube simulation variant was one that produced, on average, about 500 MB per fp32 TFLOP-hour of compute. The jobs were configured to use, on average, about 5 fp32 TFLOP-hours of compute each, thus each producing about 2.5 GB of output.

*A. Validating the setup*

Given the significant expense of a full-scale Cloud burst, we first executed a smaller Cloud burst, with the aim of both validating the expected runtime and data volume characteristics of the workload, and the feasibility of the expected peak data throughput.

For the feasibility run, we provisioned only one dedicated network link from each on-prem storage location into each of the Cloud providers, targeting the biggest compute region in each. We then provisioned a modest number of compute resources in a very short amount of time, around 10 fp32 PFLOPS in each region, and let them run for a few hours. This allowed us to reach peaks of about 7 GBps to UW storage and 2 GBps to UCSD storage, as shown in Fig. 2. As expected, the network traffic was very spiky, resulting in an average network throughput of less than half that.

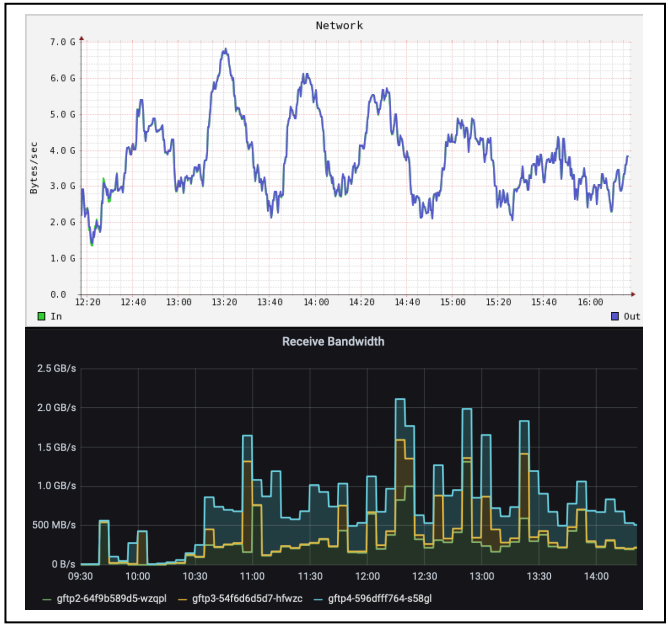

Fig. 2. Screenshots of network monitoring tools during the height of the validation run. Above: UW storage networking. Below: UCSD storage networking.

The single network link throughput pattern was similarly spiky for over-provisioned ones. We had however requested too many compute resources associated with one of the links, resulting in heavy congestion, as seen in Fig. 3. As expected, the data transfer times drastically increased on that link, and for a fraction of jobs even failed due to timeouts, resulting in significant waste of compute resources. During the validation run we identified the root cause of the problem only after the fact, which was unfortunate, but this led us to implement additional safeguards for the full-scale Cloud burst.

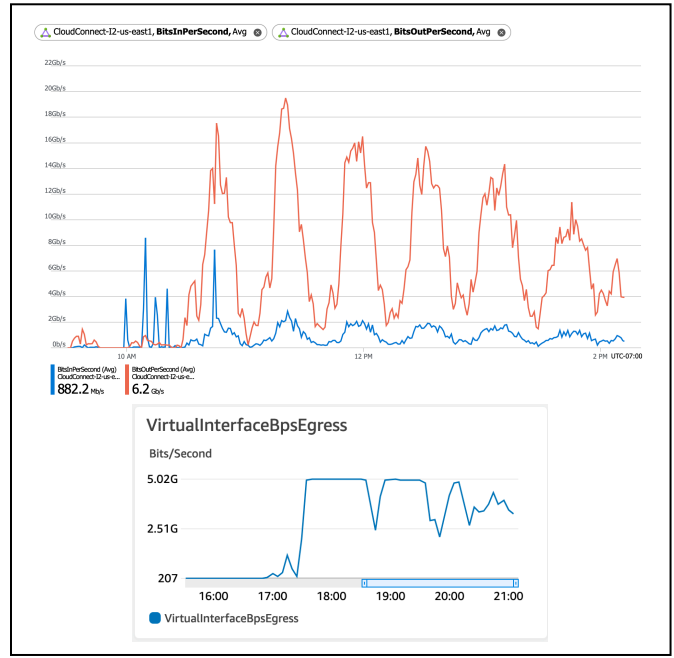

Fig. 3. Screenshots of network monitoring tools during the height of the validation run. Above: An over-provisioned dedicated network link. Below: A saturated dedicated network link.

We sustained the run for about 4 hours, and then rapidly de-provisioned the resources. During that time, about 24k jobs completed, giving us a reasonable statistic to validate the runtime and data output expectations. As expected, the mean data output size was about 2.4 GB, while the runtimes mostly correlated with the theoretical fp32 FLOPS of the various GPUs in use, as seen in Table III.

TABLE III. SUMMARY OF NETWORK EGRESS COSTS IN THE USA

| GPU Type (NVIDIA) | Mean runtime (seconds) | Theoretical fp32 TFLOPS |
| --- | --- | --- |
| T4 | 2350 | 8.1 |
| P100 | 2100 | 9.3 |
| P40 | 1950 | 11.8 |
| V100-PCIe | 1800 | 14.0 |
| V100-SXM2 | 1700 | 14.9 |
| A100-SXM4 | 1200 | 19.5 |

The validation run also allowed us to fully appreciate the amount of time needed to provision multiple dedicated links at a time; it took several hours to provision the 5 links used. Many of the steps could potentially be further automated, but it would require a non-trivial amount of development work. Some bugs in the Internet2 Cloud Connect services were also discovered. While the discovered bugs were eventually fixed, it left us with the impression that we could not rely on having all the desired links provisioned only hours before any large-scale Cloud burst.

## B. The main Cloud burst

The full-scale Cloud burst was launched about a week after the validation run, on Wed Nov. 4[th] 2020. This allowed us to add additional control and monitoring procedures in place with the aim to minimize inefficiencies. The provisioning of the dedicated network links started in the morning of the day before and was completed the morning of the Cloud burst. This was driven partially by caution and partially by the 2-hour time zone difference between the three main people involved; as mentioned in the previous section, dedicated network link provisioning is a tandem activity between the Cloud user and on-prem networking personnel. In total, we provisioned 21 dedicated network links, as described in Table II. Whenever possible, we provisioned enough network capacity to serve 150% of the average egress traffic at forecasted peak compute capacity. Unfortunately, Internet2 did not have enough peering capacity to reach that objective for the largest AWS and GCP regions, so we provisioned the maximum amount that was available there.

The Cloud compute provisioning pace was intentionally slow, in order to spread out the job startup times, and consequently the job network activity periods. Moreover, we gave preference to the most cost-effective compute instance types available in each of the Cloud regions, especially in regions where we expected to be network limited. After a ramp up of about 2 hours, we sustained a comfortable plateau of about 80 fp32 PFLOPS for slightly over an hour and then final pushed the provisioned compute capacity to 100 fp32 PFLOPS for another hour, as shown in Fig. 4. The de-provisioning phase was significantly faster, driven mostly by already-running job termination spread.

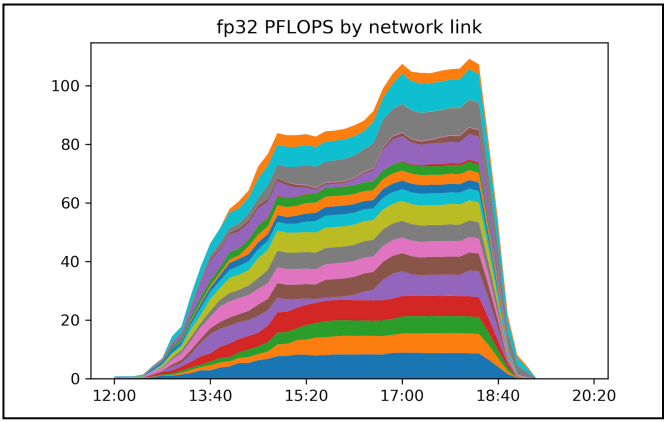

Fig. 4. Amount of provisioned Cloud compute capacity in fp32 PFLOPS over time, Pacific time zone. Each color represents a distinct dedicated network link.

During the whole period we monitored both the aggregated network throughput and per-link network utilization. Following the initial wavy behavior during ramp up, both aggregated network throughput and larger individual dedicated network link throughputs mostly stabilized, as seen in a couple select screenshots in Fig. 5. In the same figure you can also see that the UW research network was almost saturated at peak, exceeding 90 Gbps; while we were not responsible for all of it, at least 80 Gbps, i.e. 10 GBps, can be attributed to our Cloud run. Peaks of about 2 GBps were additionally flowing into UCSD.

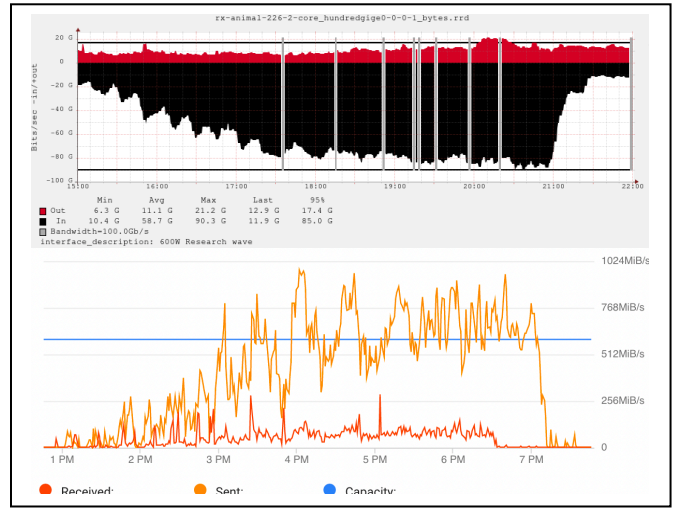

Fig. 5. Screenshots of two network monitoring tools during the main Cloud run. Above: UW reseach network link. Below: One of the provisioned dedicated Cloud network links.

The compute capacity forecast however turned out to be very inaccurate, resulting in several under-utilized dedicated network links. Moreover, we reached a plateau of cost-effective compute instances sooner than expected, which led us to spread the provisioning requests over more compute instance types than initially planned in some regions. Those compute resources were however significantly faster than the cost-effective ones, which we failed to properly compensate for, resulting in slight over-provisioning of compute resources associated with a few links. This resulted in network congestion and significantly longer output data transfer times there, as seen in Fig. 6.

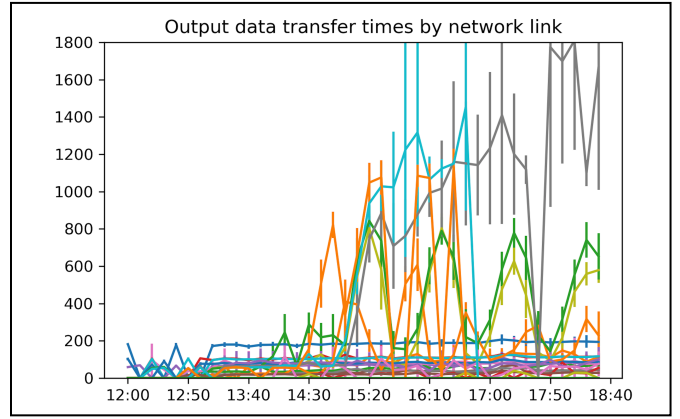

Fig. 6. Network transfer times of output files during the main Cloud run. Mean and standard deviation over time, using 10 minute bins. While upload times for jobs associated with most network links were very low, as expected, jobs associated with the few congested links spent an unacceptable amount of time transerring files.

While the abovementioned issues were unfortunate, the Cloud run overall was still a success. The integrated 225 fp32 PFLOP-hours of compute resulted in 130 TB of output data being delivered to on-prem storage, as shown in Fig. 7. Out of those, 15 GB went to UCSD and the rest to UW storage. This data volume was spread over about 54,000 output files.

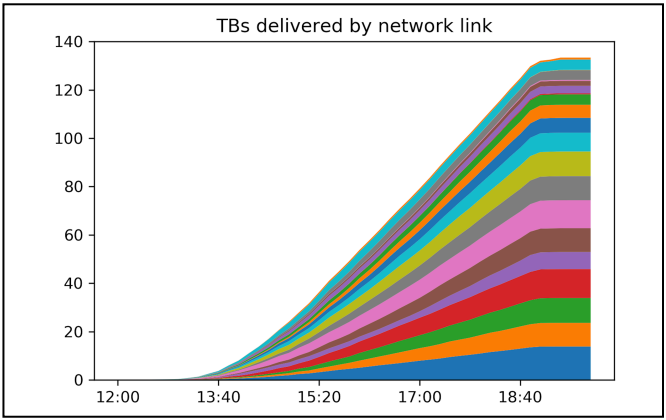

Fig. 7. Amount of data delivered to on-prem storage during the main Cloud run. Each color represents a distinct dedicated network link.

Nevertheless, the operational experience was less than optimal. Trying to provision unpredictable compute capacity, as is the case with preemptible resources, to maximize the provisioned dedicated network capacity in real time is both hard and frustrating. While we showed it is doable, within limits, we would not recommend it as a standard practice. We plan in the near future to explore intermediate output buffering mechanisms to simplify operations and improve efficiencies.

## VI. Cost analysis

The total cost incurred on Nov. 4$^{th}$ was approximately $31,000, all included, with about $5,500 of that being spent on networking.

As mentioned in the previous section, the total egress data volume was about 130 TB, so the effective metered cost was about $42/TB. If we had not used dedicated network links, the list price of metered egress data would have been about $83/TB, giving an estimated total of about $11,000 for that amount. The added effort of provisioning the links and pairing them with Cloud compute resources thus paid off in the form of almost 50% lower price.

To further put networking costs in perspective, Table IV contains the estimated costs of preemptible compute and networking on a per-instance type basis, using the GPU types as the discriminator; estimated prices were valid as of December 2020. As can be seen, for the application run in this Cloud run, networking costs would have exceeded compute costs on cost-effective compute instances, if we did not use the dedicated network links.

TABLE IV. Average cost for a single IceCube job, grouped by GPU type

| GPU Type (NVIDIA) | Compute cost per job | Network cost per job | |
|---|---|---|---|
| | | Dedicated link | Default route |
| T4 | $0.12 | $0.08 | $0.16 |
| P100 | $0.23 | $0.08 | $0.16 |
| P40 | $0.27 | $0.08 | $0.16 |
| V100-PCIe | $0.17 | $0.08 | $0.16 |
| V100-SXM2 | $0.40 | $0.08 | $0.16 |

The total relative cost savings are less impressive, since there was surprisingly little cost-effective Cloud compute capacity available that day; only about 6% of jobs ran on NVIDIA T4 GPUs and another 3% ran on NVIDIA V100-PCIe GPUs. About 40% of all the jobs ran on the least cost-effective NVIDIA V100-SXM2 GPU-providing instances.

Finally, we would like to emphasize that this was an experimental, one-of-a-kind setup and thus we are not providing an estimate of the amount of human effort needed, as it would not be representative of a more routine setup.

## VII. Conclusion

This paper describes the experience of running a half-day Cloud burst in support of IceCube production compute, that required a peak egress network traffic of over 10 GBps to return results of simulation compute generated using about 100 fp32 FLOPS of Cloud compute power. In order to minimize network related costs, all egress traffic was routed over dedicated links, which were provisioned using the Internet2 Cloud Connect service.

Overall, the Cloud run was quite a success. We integrated 225 fp32 PFLOP-hours of compute and produced 54k output files, for a total volume of about 130 TB. The total cost of networking was about $5,500, approximately 50% of the list price for egress going to the internet. The added effort of provisioning the links and pairing them with Cloud compute resources thus paid off.

That said, the operational experience was less than optimal. Trying to provision unpredictable compute capacity, as is the case with preemptible resources, to maximize the provisioned dedicated network capacity in real time is both hard and frustrating. While we showed it is doable, within limits, we would not recommend it as a standard practice. We plan in the near future to explore intermediate output buffering mechanisms to simplify operations and improve efficiencies.

Furthermore, the dedicated networking link provisioning through the Internet2 Cloud Connect service was significantly more labor intensive than envisioned, and also required extensive coordination with several independent operator groups. Development of more automation in this area would be highly desired, if such Cloud exercises were to become more routine.

### Acknowledgment

This work was partially funded by the US National Science Foundation (NSF) though grants OAC-1941481, MPS-1148698, OAC-1841530, OAC-1826967 and OPP-1600823.